# A Review of Particle Detectors for Space-Borne Self-Adaptive Fault Tolerant Systems

Marko Andjelkovic[1], Junchao Chen[1], Aleksandar Simevski[1], Zoran Stamenkovic[1], Milos Krstic[1, 2], Rolf Kraemer[1, 3]
[1] IHP – Leibniz-Institut für innovative Mikroelektronik, Frankfurt Oder, Germany
[2] University of Potsdam, Potsdam, Germany
[3] Brandenburg University of Technology, Cottbus, Germany
{andjelkovic, chen, simevski, stamenko, krstic, kraemer}@ihp-microelectronics.com

*Abstract*—The soft error rate (SER) of integrated circuits (ICs) operating in space environment may vary by several orders of magnitude due to the variable intensity of radiation exposure. To ensure the radiation hardness without compromising the system performance, it is necessary to implement the dynamic hardening mechanisms which can be activated under the critical radiation exposure. Such operating scenario requires the real-time detection of energetic particles responsible for the soft errors. Although numerous particle detection solutions have been reported, very few works address the on-chip particle detectors suited for the self-adaptive fault tolerant microprocessor systems for space missions. This work reviews the state-of-the-art particle detectors, with emphasis on two solutions for the self-adaptive systems: particle detector based on embedded SRAM and particle detector based on pulse stretching inverters.

## I. INTRODUCTION

The soft errors represent one of the most critical sources of failures in integrated circuits (ICs) employed in space missions. They are manifested as bit flips in storage elements (flip-flops, latches and SRAM cells), also known as Single Event Upsets (SEUs). These events occur when an energetic particle hits a storage element and deposits sufficient charge to alter the stored logic value. Alternatively, the particle-induced voltage glitch in combinational logic, known as Single Event Transient (SET), can cause a soft error if it propagates through the logic path and is eventually captured by a storage element.

As a result of Solar Particle Events (SPEs), the Soft Error Rate (SER) of an IC, i.e. the number of soft errors induced in a given time interval, can increase by several orders of magnitude [1]. Along with the SER variation due to radiation exposure in space, the downscaling of CMOS technologies has led to the exponential increase of the system SER [2]. This is primarily the result of dramatic increase in the number of on-chip elements with every new technology generation. Although the memory and sequential elements are dominant contributors to the overall SER because they occupy the largest area of a complex IC, the impact of combinational logic has increased significantly with the operating frequencies in the GHz range and the decrease of supply voltage and logic depth [3]. Therefore, the design of ICs for space applications requires special measures to mitigate the soft errors, i.e. to minimize the overall SER.

Besides the need for radiation hardness, the low power consumption is also an essential design requirement for the space-borne electronics, because the energy resources in space are very limited. However, the reliability requirements are usually in conflict with the power consumption constraints. For example, the reduction of the supply voltage decreases the power consumption, but increases the system SER. Moreover, the fault tolerance is traditionally based on the hardware redundancy, which increases the power consumption. Thus, the trade-off between fault tolerance and power consumption is a major goal in the design process. A cost-effective approach to accomplish this is through the self-adaptive functionality - by adapting the operating modes of the system to the application and environmental conditions [4 – 7].

A typical example of a self-adaptive system is a multi- or many-core processor. The multiprocessing platforms have been introduced to overcome the processing limitations caused by the saturation of clock frequency with the technology scaling. In the past few years, the multi- and many-core systems have gained increased interest for space missions due to the increasing demand for the on-board real-time data processing [4]. By coupling the processing cores into various configurations, the trade-off between performance, power consumption and fault tolerance can be maintained dynamically, thus extending the lifetime of the system. Depending of the radiation intensity in space and the application requirements, the fault tolerance mechanisms such as supply voltage and frequency scaling, Dual Modular Redundancy (DMR) and Triple Modular Redundancy (TMR) can be implemented [4, 5].

It is necessary to monitor the radiation level during run-time to allow for dynamic configuration of the fault-tolerance mechanisms. This is performed with the specially designed particle detectors, which operate on the principle of detecting the induced SETs or SEUs. Various types of semiconductor-based particle detectors for space applications exist and can be grouped into five main classes: (i) current detectors [8 – 15], (ii) acoustic wave detectors [16, 17], (iii) diode-based detectors [18 – 22], (iv) SRAM-based detectors [23 – 30], and (v) 3D NAND flash detectors [31 – 33]. Each type of particle detector has advantages as well as disadvantages, which are discussed in more detail in the following Sections.

The particle detectors for self-adaptive fault tolerant systems must satisfy several requirements. First, the detectors should be sensitive to a wide range of particle energies and provide the information on the particle flux, since the system SER increases linearly with the flux [34]. It is also important to monitor the variation of particle's Linear Energy Transfer (LET), because higher LET may result in multiple SEUs and longer SETs, and consequently in higher SER. The detectors should have fast response (low latency) and be robust to false alarms generated by other noise sources. Furthermore, the detectors should be integrated in the target chip to enable the in-situ monitoring of radiation exposure, and the readout electronics should introduce as low area and power overhead as possible.

However, none of the reported particle detectors [8 – 33] can satisfy all aforementioned requirements. Hence, there is a strong need for alternative solutions which can provide low-cost but accurate on-chip particle detection. Motivated by these goals, we have proposed two particle detection solutions: (i) a particle detector based on embedded SRAM [35] and (ii) a particle detector based on custom-sized pulse stretching inverter chains [36, 37]. The preliminary evaluation has confirmed that both proposed solutions offer promising advantages over the state-of-the-art particle detectors in terms of the requirements for the self-adaptive fault tolerance systems.

The rest of the paper is organized as follows. In Section II, the state-of-the-art particle detectors are briefly described. A concept of particle detection with embedded SRAM is presented in Section III, and in Section IV the particle detection with the pulse stretching inverters is discussed. The comparison of the proposed solutions with the existing ones, in terms of the main performance metrics, is given in Section V. In Section VI, an example of a self-adaptive fault tolerant mutiprocessing system with particle detection is presented. The main directions for future work are outlined in Section VII.

## II. STATE-OF-THE-ART PARTICLE DETECTORS

### A. Current Detectors

As energetic particles induce the current pulses in the target semiconductor device, the use of current sensors is a common approach for detecting these events. A simple design of a current sensor for detection of energetic particles was proposed in [8]. This sensor was connected to the power supply rail of SRAM. However, it was not suitable for detection of particle strikes in combinational logic because of the difficulty to differentiate the signal induced by a particle from the normal signal. An improved current sensor, known as the Built-in Bulk Current Sensor (BBICS), was proposed in [9, 10]. Instead of connecting to supply rail, BBICSs are connected to the bulk terminal of respective transistors. Separate BBICSs are needed for PMOS and NMOS transistors. When the bulk current exceeds the threshold level, a flag signal is generated by the sensor. The simplest structure of BBICS is composed of three transistors, as illustrated in Figure 1, but more sophisticated versions are more precise and reliable [11 – 13].

The major advantage of BBICSs is that they can provide the information on the faulty location, since they are connected directly to the target circuit. This enables to activate the error correction mechanisms only within the affected subcircuit. It is not necessary to connect a sensor to each transistor, but one sensor can be utilized to monitor tens or thousands of transistors [12, 13]. This can be used as a guideline in planning the number and spatial distribution of BBICS on a chip, in order to achieve high detection efficiency with optimal number of sensors.

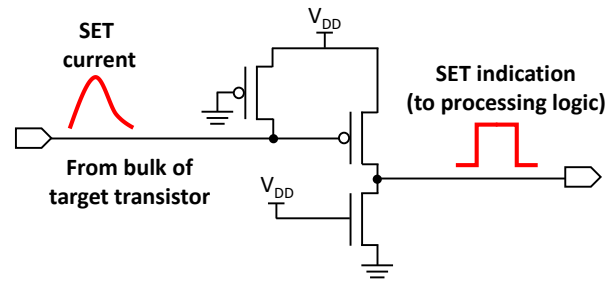

Figure 1: A simple BBICS design [10]

Nevertheless, the application of BBICS is associated with certain limitations. The key disadvantage of reported BBICS implementations is that only the particle strikes in the target circuit can be detected, but the information on the particle flux cannot be obtained directly. As the BBICSs are distributed across the chip, it is necessary to implement additional logic for collecting the data from all sensors and calculating the error rate from which the particle flux can be determined. However, there are no reports on any such implementations. In addition, the laser experiments performed on one version of the current detector [14] have revealed that BBICS sensitivity deteriorates with the increasing number of monitored transistors. A possible improvement by using the triple-well CMOS has been proposed [15], but this is not applicable to technologies with one or two wells. Moreover, as the BBICSs are connected to the target logic, they may be prone to other noise sources (e.g. substrate noise), which could lead to the triggering of false alarms.

### B. Acoustic Wave Detectors

The monitoring of soft errors with acoustic wave detectors has been proposed in [16, 17]. Namely, a particle strike can generate the intense shock (acoustic) wave, which propagates through the substrate of the target circuit. For detecting such waves, a cantilever-like structure as depicted in Figure 2 can be used [17]. It basically acts as a capacitor, and the particle strikes can be detected by measuring the change of the capacitance of the gap in the cantilever. For this purpose, the mixed-signal processing logic is needed.

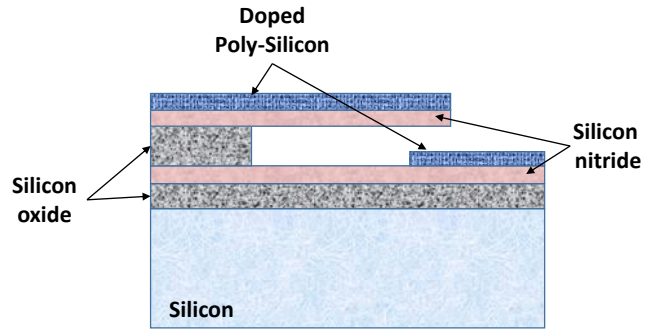

Figure 2: Cross-section of cantilever structure [17]

The solution proposed in [17] can be fabricated in CMOS technology, allowing easy integration into standard ICs. A single cantilever structure occupies the area of around 1 μm$^2$, which is roughly the area of an SRAM cell in 45 nm CMOS technology [17]. In order to achieve a sufficiently large sensing area, a mesh implementation of multiple detectors is required. Proper dimensioning of the acoustic wave detector and choice of the appropriate number of detectors for the target chip is essential for achieving high sensitivity to particle strikes. Detailed guidelines for choosing the detector dimensions and for calibrating the detector are given in [17].

Similarly to BBICS, the acoustic wave detectors enable to detect the exact location of soft errors. This is achieved based on the relative time difference of arrival of acoustic wave for different detectors, using the algorithm given in [17]. However, the main drawback of the acoustic wave detectors is that their functionality has still not been verified in practice. As these detectors need to be distributed across the chip, like the BBICS, they can provide the local detection of particle strikes, but for measuring the particle flux and LET is necessary to employ more complex processing circuitry.

*C. Diode Detectors*

The *p-n* junction (diode-based) detectors are one of the most widely used types of particle detectors. They are available in various implementations such as strip detectors, active pixel detectors and scintillator-coupled photodiodes [18 – 22]. In all implementations, the detectors are operated in reverse bias to achieve the minimum leakage current and maximum depletion layer width, thus ensuring the high detection efficiency. Radiation can induce either continuous or pulsed current pulse in the detector, depending on the radiation intensity. Measuring the induced current enables to acquire the complete information on the radiation exposure and determine with high accuracy the induced charge, particle LET, flux, and energy spectra.

However, the use of diode-based detectors for the purpose of triggering the dynamic fault tolerance mechanisms in a target IC may be too costly because different technologies have to be combined. The diodes are usually not manufactured in the same technology as the target system, which makes it challenging to integrate them on the same chip. Moreover, the need for mixed-signal processing increases the overall cost and complexity of the system. A typical structure of processing logic for a single diode is composed of a preamplifier, a pulse shaper, an analog-to-digital converter and a processor, as illustrated in Figure 3. As the practical implementations may be composed of hundreds or thousands of diodes on the same substrate, the hardware and power overhead due to the processing logic may be too high.

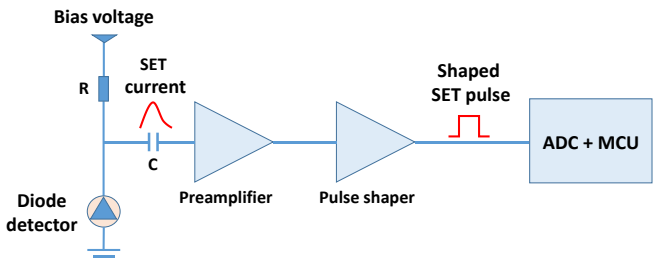

Figure 3: A processing channel for diode-based detector

*D. SRAM Detectors*

The use of commercial or custom-designed SRAMs as particle detectors, implemented as stand-alone ICs, has proven to be a very useful solution for soft error monitoring in various terrestrial and space applications [23 – 30]. The operation principle is based on counting the number of particle-induced SEUs in SRAM cells. When a particle hits a sensitive transistor within the cell, and deposits the energy exceeding the critical charge, the respective logic state will be changed from 0 to 1 or vice-versa. In general, the sensitivity is proportional to the size of SRAM (number of SRAM cells). The most common implementations employ the six-transistor (6T) SRAM cells as illustrated in Figure 4. Based on the detected number of SEUs and the cross-section of SRAM obtained experimentally, the particle flux can be calculated. The SEUs are detected and corrected using some of the well-known Error Detection and Correction (EDAC) mechanisms and memory scrubbing. The response time of the SRAM-based detector is determined by the scrubbing rate, which is on the other hand defined by the clock frequency of the system.

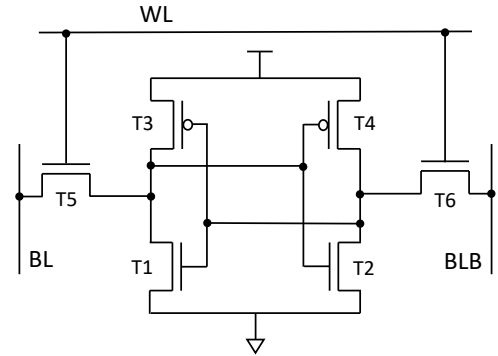

Figure 4: 6T SRAM cell

The main advantage of SRAM detectors is simple operating principle, no need for analog processing and possibility of manufacturing in the same technology as the standard ICs. However, this approach has several limitations. In the stand-alone implementations [23 – 30], the area overhead due to the EDAC logic may be too large. From functional point of view, the EDAC techniques suffer from the limitation in the number of detectable and correctable errors, which may lead to the error accumulation as a result of multiple upsets. Moreover, due to scrubbing the response of SRAM detectors may be slower compared to other solutions.

While most reported SRAM-based solutions have been used only for flux measurement, it is also possible to measure LET with custom-designed SRAM detectors. A solution presented in [29, 30] uses the custom-designed SRAM detector which generates SET pulses in response to particle strikes, and the analog processing logic is used to amplify the pulses and then measure their amplitude. Based on measured pulse amplitude, the energy and LET of incident particles can be determined. In this case, all cells are connected in parallel, so that a particle strike induces a pulse which can propagate to the output. The solution is designed as a standalone spectrometer and as such is not suitable for the on-chip integration.

*E. 3D NAND Flash Detectors*

Recently, the use of 3D NAND flash memory with floating gate transistors as a heavy-ion detector has been proposed [31]. Although the detectors based on floating gate transistors have been used for total dose measurement [32, 33], the work [31] is the first to verify the applicability of this concept for detection of energetic particles. The operating principle relies on measurement of the threshold voltage shift of floating gate transistors due to the charge deposited by the incident particles. This allows not only to measure the error rate, but also the particle LET. In addition, due to the 3D structure of memory, the angle of incidence can be estimated. Furthermore, the 3D structure allows to differentiate between the errors induced by incident particles and the errors due to electrical noise, as well as to differentiate between single and multiple upsets.

Although the 3D NAND flash with floating gate transistors is a promising solution with substantial benefits over other detectors such as SRAM, the main limitation currently is difficulty in integrating it in the target chip. Due to the 3D structure and floating gate technology, this approach may be too complex for integration into a conventional planar CMOS IC designed with standard design tools. In addition, the processing electronics may be complex and costly because it is necessary to measure precisely the change of the threshold voltage of floating gate transistors, which requires the use of analog processing circuitry and analog-to-digital converters.

### III. EMBEDDED SRAM AS A PARTICLE DETECTOR

As alternative to the conventional stand-alone SRAM-based particle detectors described in previous Section, we have proposed the use of embedded SRAM as a particle detector [35]. The idea is to employ the standard on-chip SRAM memory as a particle detector in parallel to its normal data storage function. The detection principle is the same as for the stand-alone SRAM detectors discussed in previous Section, i.e. the SEUs detected in SRAM cells are counted and from this information the particle flux can be determined. A similar approach, based on Block Random Access Memory (BRAM) in FPGA was introduced in [1]. However, in contrast to all previous solutions, the proposed embedded SRAM monitor incurs significantly less area overhead because the available on-chip resources are utilized for particle detection. An important feature of the proposed solution is the capability to detect the permanent faults in SRAM cells. This is essential for maintaining the accurate SEU measurements in long-term missions, where the permanent errors occur due to the gradual device wear-out.

Figure 5 shows the block diagram of the embedded SRAM with the support for particle detection. It consists of a Synchronous SRAM (SSRAM) with five 512k × 8-bit asynchronous SRAM blocks, a Control Unit, a Scrubbing module and an EDAC module. Four memory blocks are used for data storage and particle detection, while one block is allocated for storing the 7-bit EDAC syndrome computed on each 32-bit word written in the other four memory blocks. Thus, the user sees effectively a 16-Mbit memory device. The memory blocks are based on the 6T SRAM cell shown in Figure 4. Each read, write or scrubbing cycle uses the EDAC module and involves the access to 32 bits selected by a 19-bit address. As the sequential logic in the Control Unit, EDAC and Scrubbing modules is inherently sensitive to SEUs, the Triple Modular Redundancy is applied to all flip-flops.

The functions of EDAC and Scrubbing modules is to protect the memory cells against SEUs and detect the single and double bit errors as well as permanent faults in each memory word. The built-in EDAC module performs the Single-Error Correction and Double-Error Detection (SEC-DED) with (39, 32) HSIAO code. The HSIAO code was chosen because it provides fast coding/decoding with low hardware overhead. The three 8-bit Error Counters are integrated in the Control Unit to count the single and double bit errors and permanent faults. Any error that cannot be corrected by EDAC is considered as a permanent error. A detailed description of the algorithm for detection of single, double and permanent faults can be found in [35]. The number of detected faults is stored in the Register File to avoid duplicate counting of the double and permanent faults.

The primary role of the Scrubbing module is to avoid accumulation of radiation-induced soft errors. The scrubbing module periodically reads the memory locations when the chip is in the idle state. When the error is detected, the EDAC procedure is activated. The scrubbing is entirely autonomous and transparent for the user, which means that the user can access the SSRAM even if the scrubbing procedure is in progress. The scrubbing rate (response time) can be configured by the user, but it is limited by the operating frequency.

The proposed SRAM monitor has been designed in IHP's 130 nm bulk CMOS technology with the nominal supply voltage of 1.2 V. The recommended operating frequency for this design is 50 MHz. For this frequency, the minimum scrubbing rate is 42 ms. With respect to the total area of the design, the introduced area overhead is less than 1%, while the power overhead is even less, below 0.1 %. The elements contributing to the area and power overhead are the Error Counters and the Register File, while all other hardware resources are employed in any rad-hard SRAM. A detailed discussion of the synthesis results can be found in [33].

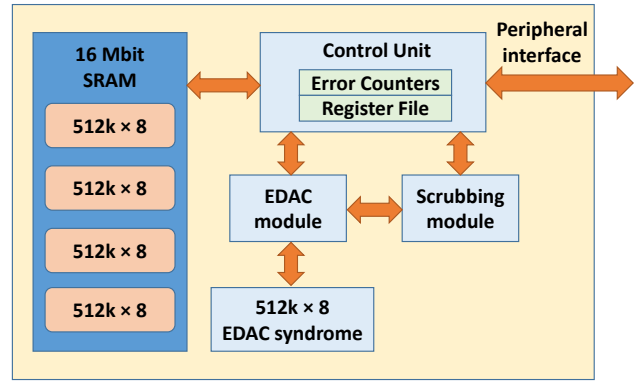

Figure 5: Embedded SRAM as a particle detector

However, the main issue with the on-chip SRAM used as a particle detector is the limited sensitive area. In general, the SRAM should be as large as possible to obtain sufficiently large sensitive area and thus ensure the high probability of particle detection. Previous studies have shown that the SRAM with the capacity of several Gbit is needed for sufficient sensitivity, but these solutions are based on standalone SRAMs. When the data-storage SRAM within the target chip is used as a particle detector, its sensitive area will be constrained by the application requirements. The size of the on-chip SRAM is usually limited to tens or hundreds of Mbits. To acquire statistically relevant number of SEUs with a smaller on-chip SRAM, it is necessary to employ longer detection intervals.

## IV. PULSE STRETCHING INVERTERS AS A PARTICLE DETECTOR

The application of custom-sized pulse stretching inverter chains as particle detectors has been proposed in our previous work [36 – 38]. The idea is to measure the SET count rate and SET pulse width variations. Thereby, the particle flux can be determined in terms of the SET count rate, while the LET variations can be determined in terms of the SET pulse width variations. It is important to note that this solution cannot measure the exact SET pulse width, but only to sort the detected SET widths into several distinct ranges. This is due to the fact that digital processing logic, as a simple and low-cost alternative to the analog processing used for diode detectors, has been chosen in this case. Nevertheless, the information on the SET count rate and SET pulse width variation is sufficient for the target self-adaptive fault tolerant systems.

A basic sensing element consists of two inverters connected in series, and this configuration is denoted as a Pulse Stretching Cell (PSC). By setting the fixed logic level at the input of PSC, two transistors will always be in on-state while the other two will be in off-state. The off-state transistors are sensitive to particle strikes, while the on-state transistors act as restoring elements (provide the current to compensate the particle-induced charge). The PSCs have skewed sizing, i.e. in one inverter the PMOS transistor has larger channel width than NMOS transistor, while in the other inverter the NMOS transistor has larger channel width than the PMOS transistor. To achieve sufficiently large sensing area of a PSC and thus increase the probability of particle strikes, the off-state transistors should have as large channel width as possible. On the other hand, to decrease the restoring current and thus increase the sensitivity, the on-state transistors should have small channel width and large channel length. Furthermore, the skewed sizing ensures that the SET pulse induced in the PSC is stretched as it propagates through the chain. Therefore, even the low energy particles can results in observable SETs. A detailed description of the transistor sizing for the PSC can be found in [36 – 38].

Using a single PSC is generally not sufficient because the two sensitive transistors still have quite small sensing area. The sensing area can be increasing by connecting an appropriate number of PSCs in series or in parallel. In serial configuration, two detector versions are possible: (i) a long chain of PSC or (ii) a number of shorter PSC chains connected by an OR tree. In parallel configuration, the number of PSC that can be connected in parallel is limited due to the loading effects, i.e. a large number of PSCs in parallel reduces the sensitivity. Thus, a number of arrays made of PSCs connected in parallel have to be employed, as illustrated in Figure 6. The serial configuration is suitable only for measuring the SET count rate because the SET pulse width changes very little over a wide range of LET. On the other hand, the parallel configuration enables to capture both the SET count rate and the SET pulse width variation.

Both configurations have been evaluated with SPICE simulations, using the bias-dependent current model to simulate SET effects. The analysis was performed for IHP's 130 nm CMOS technology. It was shown that with large off-state transistors and small on-state transistors, the threshold LET is below 0.2 MeVcm$^2$mg$^{-1}$. This is lower than the threshold LET of standard logic cells, and also lower than the LET of common particles encountered in space. For the parallel configuration depicted in Figure 6, the SET pulse width changes by approximately 550 ps in the LET range from 1 to 100 MeVcm$^2$mg$^{-1}$, whereby the maximum SET width is in the order of several ns. On the other hand, in the serial configuration the output SET pulse width is directly proportional to the number of cascaded PSCs and can be from hundreds of ps to hundreds of ns. Therefore, the pulse stretching detector is expected to have faster response than the SRAM-based detectors. Moreover, the pulse stretching detector is immune to error accumulation because of the transient nature of SET effects.

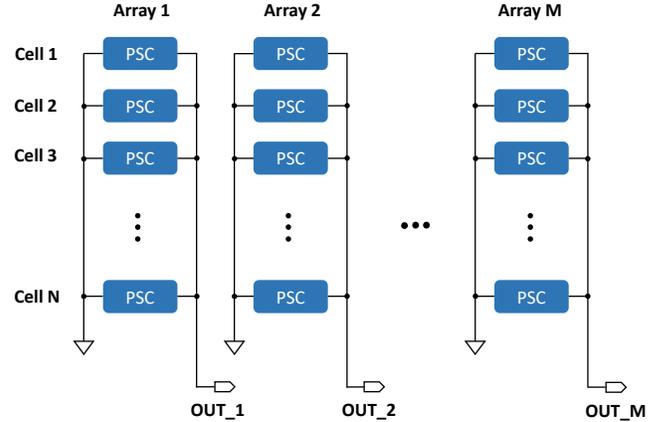

Figure 6: Particle detector based on arrays of PSCs connected in parallel

Figure 7 illustrates the general architecture of the processing logic for the particle detector composed of parallel arrays of PSCs illustrated in Figure 6. The outputs of all pulse stretching arrays are connected to a standard OR-tree to obtain a single output which is then interfaced to the processing logic. An SET induced in any array will propagate to the output of the OR-tree and then further through the SET filters and respective SET counters. The SET filters allow the propagation of SET pulses within predefined pulse width ranges. Thus, the corresponding counters store the number of detected SETs with the predefined pulse widths. The control unit reads the current state of all counters, stores the acquired results in register file and resets periodically the counters. The standard hardening measures can be applied to the processing logic.

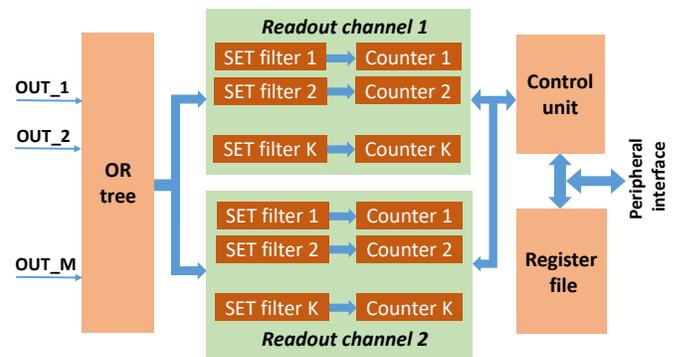

Figure 7: Readout circuit for pulse stretching particle detector based on parallel PSC arrays

## V. COMPARISON OF PARTICLE DETECTORS

Based on the published results, a comparative analysis of the discussed particle detectors in terms of six performance metrics is presented in Table 1. The advantages and advantages of each type of detector should be carefully considered in selecting the appropriate detector for a particular application. For the space applications where the self-adaptive dynamic fault tolerance is required, it is important that the particle detector is integrated in the same chip with the target system. This enables to sense directly the radiation to which the target system is exposed. In that context, the two detectors proposed in our previous work offer essential advantages over the state-of-the-art solutions.

The main advantage of embedded SRAM-based detector over all other solutions is that it serves as a standard data storage memory in a target system. This results in negligible area and power overheads since the existing on-chip resources are used for particle detection. As a result, the cost of implementation is lower compared to other solutions. In addition, the possibility of detecting the permanent errors is important advantage over all other detectors, as none of the existing solutions supports this functionality.

On the other hand, the particle detector based on the pulse stretching inverter chains offers the possibility to measure the LET variation, which is possible also with diode and 3D NAND flash detectors. However, compared to these detectors, the pulse stretching detector employs simple digital processing logic, which minimized both the area and power overheads and thus the overall cost. Moreover, the immunity to multiple errors is an advantage over the conventional SRAM detectors.

Table 1: Comparison of particle detectors

| Type of detector | Probability of false alarms | Complexity of readout logic | Hardware/power overhead | Sensitivity to multiple errors | Ability to monitor LET variation | Additional functions |
|---|---|---|---|---|---|---|
| Built-in current detector | Moderate | Low | Medium | No | No | No |
| Acoustic wave detector | Moderate | Moderate | Medium | No | No | No |
| Diode detector | Low | High | High | No | Yes | No |
| Stand-alone SRAM detector | Low | Low | High | Yes | No | No |
| 3D NAND flash detector | Low | High | High | No | Yes | No |
| Embedded SRAM detector | Low | Low | Low | Yes | No | Data storage and detection of permanent errors |
| Pulse stretching detector | Low | Low | Medium | No | Yes | No |

## VI. APPLICATION SCENARIO: SELF-ADAPTIVE QUAD-CORE PROCESSING SYSTEM

To illustrate the operation of a self-adaptive fault tolerant multi-processing system with a built-in particle monitor, we have chosen a quad-core platform as an example.

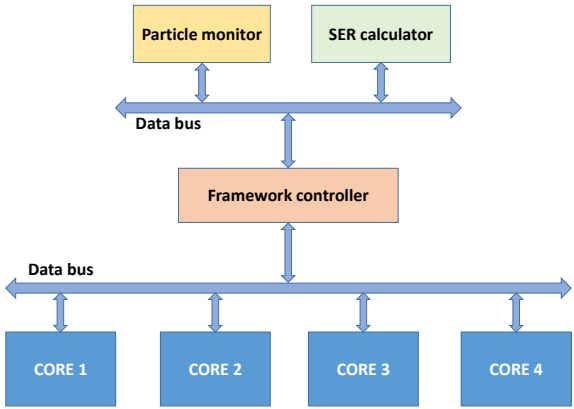

Figure 8: General architecture of a self-adaptive quad-core processing platform

The block diagram of the self-adaptive fault tolerant quad-core system is illustrated in Figure 8. The basis of the system is the *Waterbear* framework controller which enables to select the three main operating modes [4]:

- **High performance mode**: the multiprocessor operates as a common multiprocessor, i.e. each core executes its own task. This mode is selected according to the application requirements.

- **Destress (or low-power) mode**: a single core is operating while the others are clocked- or powered-off to reduce aging and save power. The anti-aging technique known as Youngest-First Round-Robin (YFRR) core gating [39] is employed to de-stress the operating core by transferring the workload to a resting core. This is done periodically and with the help of special aging monitors embedded in each core [40].

- **Fault tolerance mode**: the processing cores are coupled into various fault-tolerant configurations such that they are executing each instruction simultaneously. The voting is performed in each cycle by a voter unit which initiates the actions (e.g. interrupt request or reset) when the mismatch between the core outputs is detected [4].

During operation under radiation exposure in space, the particle monitor generates the information on the SET or SEU count rate and LET (if the chosen detectors supports the LET measurement). The SER calculator processes the information from the particle detector to determine the real-time SER variation of the multi-core system. The SER calculator can be extended with a hardware accelerator module for prediction of SPEs, at least one hour in advance, based on the supervised machine learning, as detailed in [41, 42]. This functionality allows for the early detection of the increasing radiation levels which result in increased SER, and timely activation of the respective fault tolerant mechanisms.

Based on the measured or predicted SER, various fault-tolerant solutions can be applied at the core level to achieve the radiation hardness, such as:

- **Supply voltage and frequency scaling:** By increasing the supply voltage and decreasing the operating frequency, the SER is reduced at the cost of increased power consumption and reduced processing speed. This approach can provide limited improvement in SER which could be valuable at low and medium level radiation levels. In this case, either all cores are engaged in parallel processing or some of them may be switched off.

- **Double modular redundancy (DMR):** In this mode, the four cores can be divided into two pairs of DMR cores, such that the system is essentially operating as a dual-core system with enhanced fault tolerance. This approach is useful under medium radiation exposure, but the drawback is the reduced processing speed.

- **Triple modular redundancy (TMR):** In this mode, three cores are coupled into a TMR configuration while the fourth core is powered off. Thus, the system operates as a single core with the highest level of protection under high radiation levels. The main drawback of this approach is the reduced processing speed because all cores perform the same task.

The concept illustrated in Figure 8 is flexible and can be adopted to a larger number of cores with minor modifications of original design. To accommodate the particle detector and SER calculator, the original framework controller design requires the addition of an interface for processing the data from the added modules. If the platform is applied to a many-core system, the DMR and TMR configurations can be implemented on multiple groups of processing cores. For example, in an assumed 8-core system would be possible to have two TMR blocks operating as a dual-core processor, thus achieving the high level of fault tolerance and at the same time providing enhanced performance. This concept has been verified on an 8-core 32-bit chip demonstrator designed and manufactured in IHP's 130 nm bulk CMOS technology [5].

The main benefit of the multi-core approach in terms of fault tolerance is that the inherent hardware redundancy is used as a basis for achieving the fault tolerance. The processing cores are considered as redundant only in the fault tolerant mode while in the high performance mode they are employed for multiprocessing. As a result, the area overhead is minimal and is related only to the additional logic that is needed for selecting the fault tolerant modes.

## VII. CONCLUSION AND FUTURE WORK

In this work, the comparative analysis of several solutions for detection of energetic particles responsible for soft errors in integrated circuits is presented. The comparison was performed based on the requirements for the online particle detection in the self-adaptive fault-tolerant systems for space applications. Beside the five state-of-the-art semiconductor particle detectors (diode-based, SRAM-based, bulk built-in current, acoustic wave and 3D NAND flash detectors), we have introduced the two detector concepts which have resulted from our ongoing research – the embedded SRAM-based detector and the pulse stretching detector. The comparative analysis has shown that the two proposed solutions have remarkable advantages over the existing particle detectors regarding the self-adaptive fault tolerance applications.

Future work will be directed towards experimental validation of the two proposed particle detectors. To this end, it is necessary to conduct the irradiation campaign with the detector prototypes, in order to calibrate their response and determine the optimal design specifications.

## ACKNOWLEDGMENT

This work was done in the framework of project REDOX (funded by the German Research Foundation DFG under the grant agreement No. KR 3576/29-1), and project RESCUE (funded by the European Union's Horizon 2020 research and innovation programme under the Marie Sklodowska-Curie grant agreement No. 722325).